\documentclass[conference]{IEEEtran}
\IEEEoverridecommandlockouts
\usepackage{cite}
\usepackage{amsmath,amssymb,amsfonts}
\usepackage{algorithmic}
\usepackage{graphicx}
\usepackage{textcomp}
\usepackage{xcolor}
\usepackage{subfig}
\usepackage{cite}
\usepackage{url}
\usepackage{caption}

\def\BibTeX{{\rm B\kern-.05em{\sc i\kern-.025em b}\kern-.08em
    T\kern-.1667em\lower.7ex\hbox{E}\kern-.125emX}}

\begin{document}

\title{Reusable MLOps: Reusable Deployment, Reusable Infrastructure and Hot-Swappable Machine Learning models and services}

\author{\IEEEauthorblockN{1\textsuperscript{st} Deven Panchal}
\IEEEauthorblockA{\textit{AT\&T} \\
Middletown, NJ, USA \\
devenrpanchal@gmail.com}
\and
\IEEEauthorblockN{2\textsuperscript{nd} Prafulla Verma}
\IEEEauthorblockA{\textit{AT\&T} \\
Middletown, NJ, USA \\
pv2985@att.com}
\and
\IEEEauthorblockN{3\textsuperscript{rd} Isilay Baran}
\IEEEauthorblockA{\textit{AT\&T} \\
Middletown, NJ, USA \\
ib6391@att.com}
\and
\IEEEauthorblockN{4\textsuperscript{th} Dan Musgrove}
\IEEEauthorblockA{\textit{AT\&T} \\
Middletown, NJ, USA \\
dm4812@att.com}
\and
\IEEEauthorblockN{5\textsuperscript{th} David Lu}
\IEEEauthorblockA{\textit{AT\&T} \\
Dallas, TX, USA \\
dl1971@att.com}
}

\maketitle

\begin{abstract}
Although Machine Learning model building has become increasingly accessible due to a plethora of tools, libraries and algorithms being available freely, easy operationalization of these models is still a problem. It requires considerable expertise in data engineering, software development, cloud and DevOps. It also requires planning, agreement, and vision of how the model is going to be used by the business applications once it is in production, how it is going to be continuously trained on fresh incoming data, and how and when a newer model would replace an existing model. This leads to developers and data scientists working in silos and making suboptimal decisions. It also leads to wasted time and effort. We introduce the Acumos AI platform we developed and we demonstrate some unique novel capabilities that the Acumos model runner possesses, that can help solve the above problems. We introduce a new sustainable concept in the field of AI/ML operations - called Reusable MLOps - where we reuse the existing deployment and infrastructure to serve new models by hot-swapping them without tearing down the infrastructure or the microservice, thus achieving reusable deployment and operations for AI/ML models while still having continuously trained models in production.
\end{abstract}


\section{Introduction}
Machine Learning, Deep Learning and other statistical techniques are being increasingly employed these days to solve various kinds of problems. Over the past few years, we have seen an exponential rise in the demand for machine learning expertise in various industries to solve problems that were either not easily solvable or required considerable effort to solve. This is due to the rise and easy availability of ML / DL / AI tools libraries / techniques, which has also led to ML becoming increasingly popular. This coupled with the availability of very capable training and serving i.e., compute resources like powerful CPU’s, GPU’s, TPU’s \cite{surveymlaccelerators}\cite{TPU} and various other offerings in the cloud, has made ML models a go-to solution for many problems. What remains difficult for individuals and smaller companies, however, is the operationalization of these models \cite{p3}. While data scientists and machine learning engineers are doing a great job at data cleaning, wrangling, visualization and modeling, they find it difficult to architect a working solution complete with the ML model deployed in production, standing as a service, serving predictions to business applications that may use them, and at the same time being able to do model retraining on fresh incoming data, or changing the performance and/or the behavior of the model-all without disturbing the business applications that are using the model \cite{ml-technical-debt}\cite{mlops1}\cite{mlops2}.

\section{Societal and Managerial Impacts}

AI is widely regarded as a cross-cutting transformative technology that is being rapidly adopted across industries. While AI can help bring automation, better accuracy, improvements in healthcare, agriculture, and in general an improved quality of life, and help increase economic productivity and profits to positively affect the GDP, it is a technology that should be effectively managed to avoid adverse effects. Institutions will have to consider the societal, managerial and ethical implications \cite{doi:10.1177/0008125619863436}\cite{floridibook}\cite{8058187} \cite{8114684} as they carry out AI and Digital transformation programs. Managerial impacts could include adopting, integrating or transitioning to new technologies, rethinking business processes and customer experience and managing skillsets. Adverse societal impacts like the effect on the labor force, anomie, privacy and security concerns will have to be considered. The adoption of AI is bound to grow in coming years. Hence, it becomes all the more important that such a technology be sustainable in all respects. It should be available to all, and benefit all. The cost of using it should be reduced progressively as it is increasingly deployed. The solution we discuss in this paper tries to solve the problems that we talked about earlier - the problem with silos, wasted time, cost, labor, and the unavailability of expertise in operationalizing AI/ML models and intelligent services. The Acumos platform that we have developed, helps make artificial intelligence accessible to everyone, helps operationalize AI/ML models quickly and easily while providing many novel capabilities. In this paper we will also specifically discuss how considerable time, cost and effort savings can be obtained from Reusable MLOps and its features of Reusable Deployment, Reusable Infrastructure and Hot-Swappable Machine Learning models. For AI/ML enabled IoT applications, we often encounter legal requirements surrounding data sovereignty, data location and data transfer. Acumos Federation and Licensing facilitate fine grained sharing of AI/ML models using a model marketplace solution. This is explained in \cite{p6}.

\section{Related Work}
We developed the Acumos AI platform. Acumos is an open-source project that aims to make artificial intelligence accessible to everyone i.e., democratize AI. The Acumos platform umbrella project has many projects which help to build, share, and deploy AI applications. It abstracts the infrastructure stack and components and helps to run an out-of-the-box AI environment. This helps with the problems we discussed earlier and the Data scientists, ML engineers, Software developers can all focus on their core competencies while still being able to take complex models written in a variety of languages, using a variety of toolkits to production \cite{AcumosLF}\cite{zhao2018packaging}. Further, it allows easy sharing of models between individuals, teams, departments of the same company, or between companies. This sharing of models can be monetized using the Licensing and Federation features that Acumos provides \cite{p6}\cite{Documentation-federation}\cite{Documentation-lum}\cite{devpat}. The Acumos Model Runner component, which is one of the projects under the Acumos umbrella project, is used in conjunction with the Acumos Java client tool (another project under the Acumos umbrella project) to onboard H2O models, generic Java models (any model written in Java. For e.g., say using Weka or any other package), Spark models to the Acumos AI platform.  We will talk about the Acumos Model Runner \cite{Documentation-JavaModelRunner} and see how it can be used to operationalize a continuously trained model i.e., immediately operationalize a better performing model or seamlessly operationalize a model whose behavior i.e., inputs and outputs have changed considerably. Acumos thus makes it easy to productionalize ML models and do more by creating a zero-touch AI pipelines for model retraining and model serving through the use of Design Studio \cite{Documentation-Design-Studio}\cite{p4}. The AI4EU Experiments distribution of Acumos is hosted at \cite{ai4eu-experiments}. But you can certainly stand up your own instance of Acumos.

\begin{figure}[htbp]
\centering
 {\includegraphics[width=0.4\textwidth]{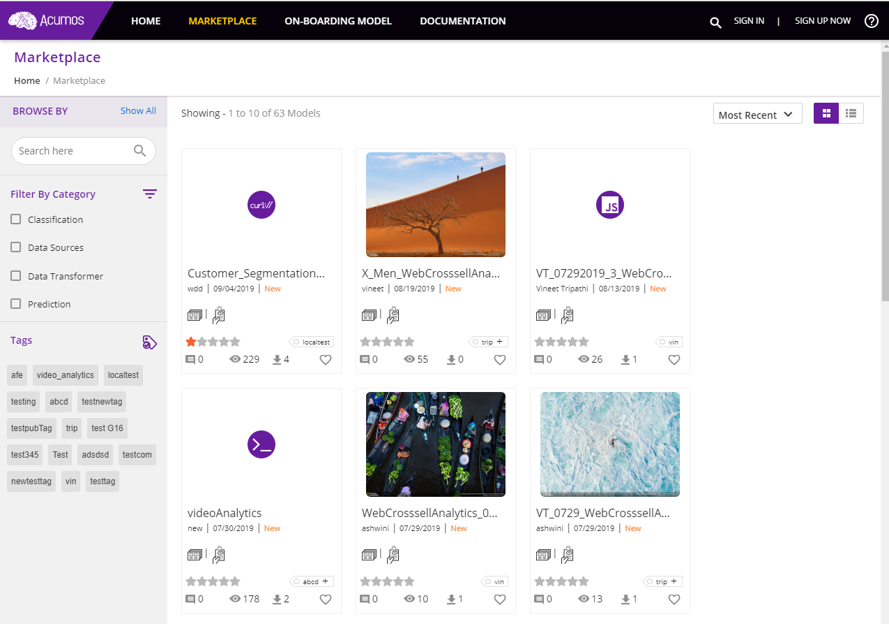}}
 \captionsetup{justification=centering}
\caption{Acumos Web Interface}
\label{Acumos-Website}
\end{figure}

\section{How a Modeler can onboard a model to Acumos using Acumos Java Client and Acumos Model Runner}
The data scientist or the ML engineer would first train a model in H2O or in Java (any framework/package) or Spark.  Depending upon the package used to create the model, the modeler can then export the model to a file (Mojo zip file for H2O, jar for Java and Spark). The modeler would then run the latest Java client tool that he can download from the Acumos Nexus repository alongside this exported model file and the Acumos Model Runner artifact that he would be able to get from the Acumos Nexus repository as well. This process would result in the Java client tool generating a bunch of artifacts (a proto file, metadata.json and modelpackage.zip) for the model. During this time, the model Runner creates a common Acumos wrapper around the ML model (irrespective of the type of model) and packages it as a microservice which exposes some REST endpoints \cite{Documentation-Javaclient}\cite{p2}.

The modeler/ data scientist could manually upload the aforementioned artifacts generated by the Java client tool into the Acumos marketplace Web GUI (Web-based onboarding) or he could have passed in some extra arguments when running the Java client tool and the client tool would have done this for him (CLI-based onboarding).

\begin{figure*}[htbp]
\centering
 {\includegraphics[width=0.9\textwidth]{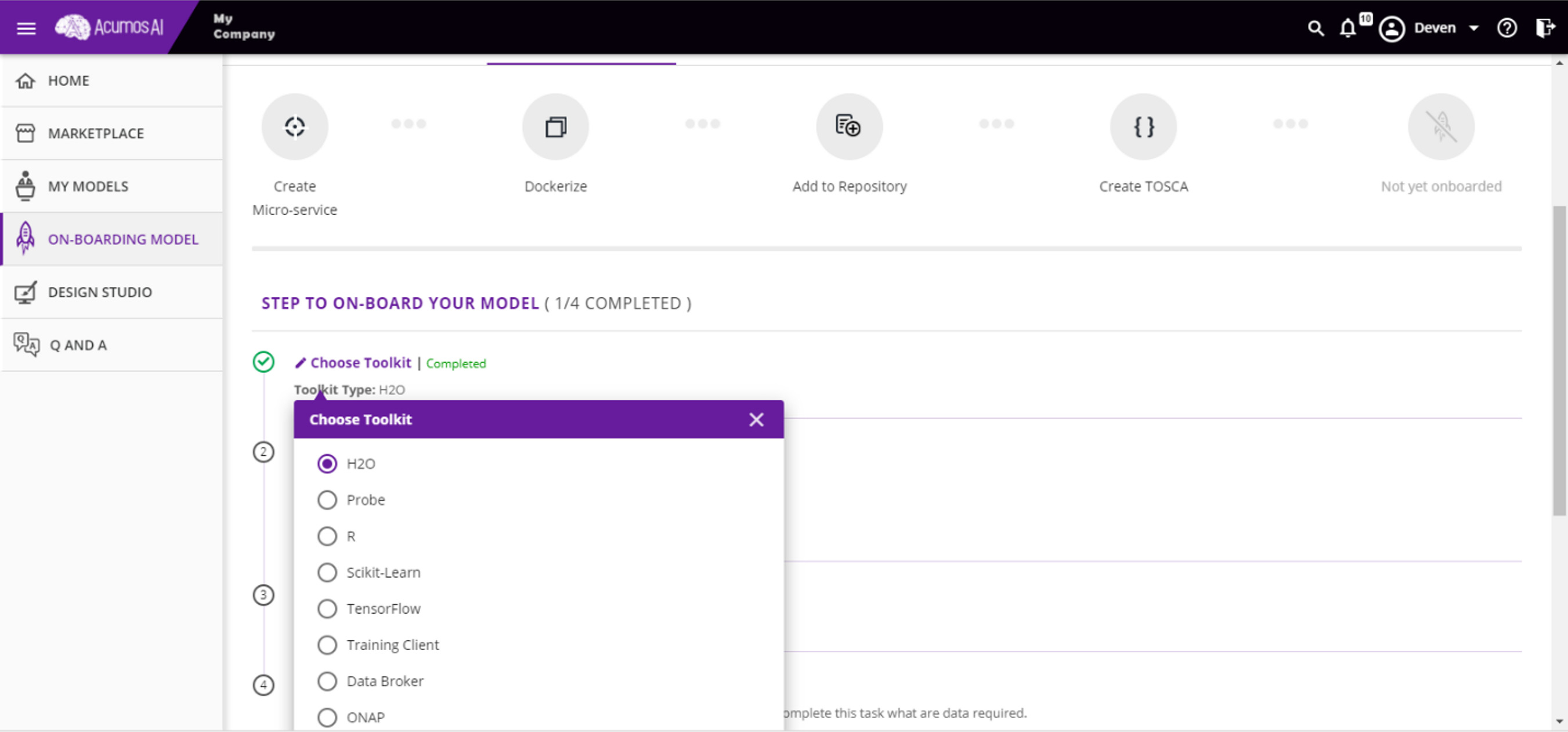}}
 \captionsetup{justification=centering}
\caption{A Machine Learning model being onboarded to Acumos }
\label{Acumos-Onboarding}
\end{figure*}

\begin{figure*}[htbp]
\centering
 {\includegraphics[width=0.9\textwidth]{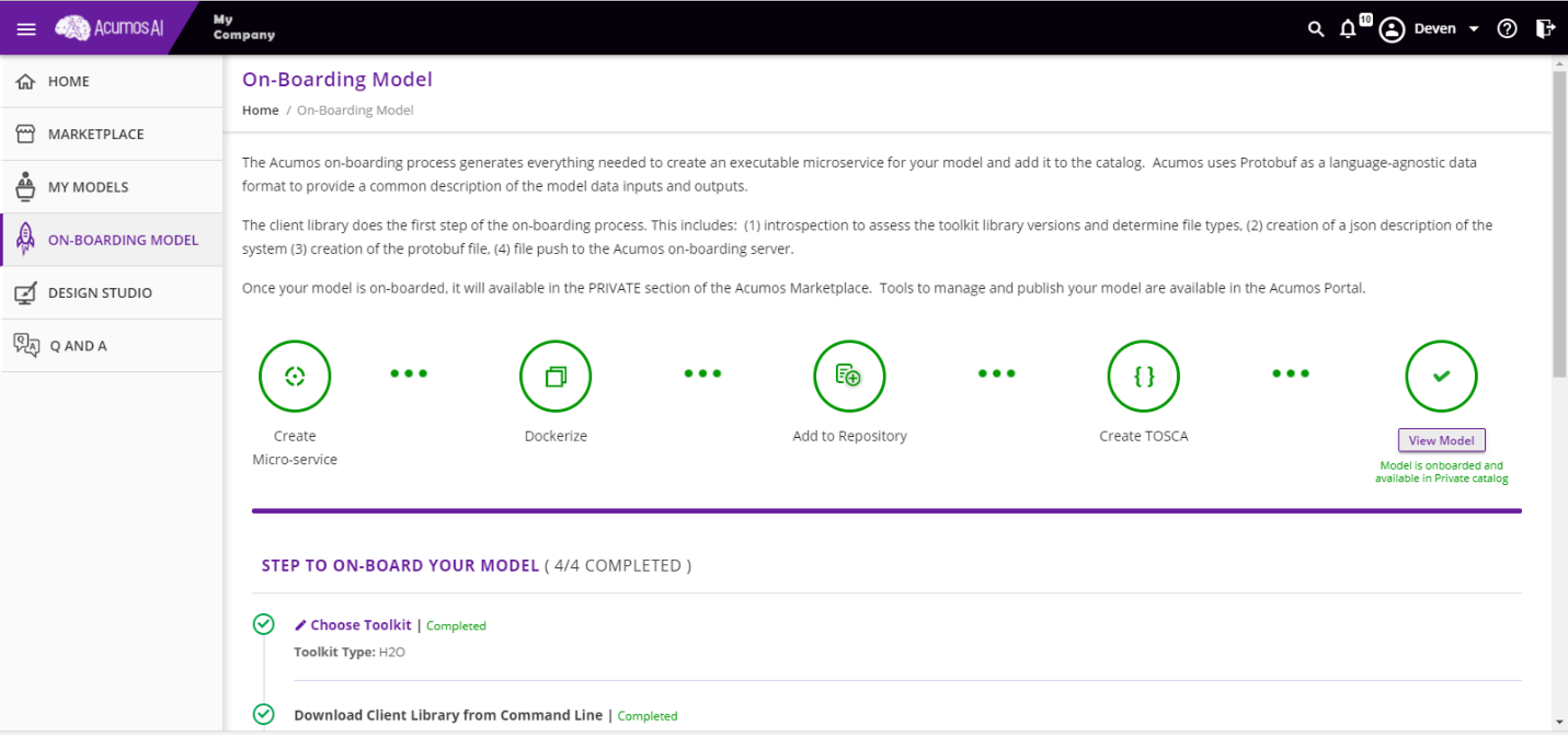}}
 \captionsetup{justification=centering}
\caption{A Machine Learning model onboarded successfully to Acumos}
\label{Acumos-Onboarding-Complete}
\end{figure*}

Once the upload/onboard to Acumos marketplace is successful, docker images and TOSCA artifacts are created. These can be deployed from within the ACUMOS platform to your favorite target like Kubernetes, AWS, Azure, GCP etc. You can also deploy it on your own docker enabled cloud/machine because ACUMOS allows you to download the docker image directly from the marketplace.  Your model is now standing as a microservice ready to serve predictions and do much more that we promised earlier. We will explore its capabilities later.

\section{Acumos Model Runner Working}
When running in conjunction with the Java client, the Generic Model runner has slightly different behavior depending on which type of model it is being used with- a Java model or a Spark model or H2o model. This is specified early on in the application.properties file. Since we do a lot of saving, retrieving, and sending (over the wire) of data into and out of the models i.e., the model microservices, Acumos have chosen to use Protocol Buffers (Protobuf) as the serialization technology. According to its own documentation \cite{protobuf}, Protobuf provides a language-neutral, platform-neutral, extensible mechanism for serializing structured data. Protobuf is well suited to applications where we need to serialize structured, record-like, typed data. It helps achieve a very compact serialized format and so has a low memory footprint and a low network footprint. It has the advantage of being fast to parse and gives so much automatically generated functionality for reading, altering, serializing, deserializing - and in general working with the underlying data.

Protobuf will help us achieve our dream of achieving a lower latency when we talk to our model microservices.

The model runner reads the input proto file, does protobuf compilation using protoc compiler and generates a Java Protobuf file. Then the model runner invokes the Java compiler to compile the Java Protobuf file to corresponding class files. These class files are then loaded into the Java Virtual Machine (JVM) dynamically at runtime using the Java classloader. These classes are then used as and when required by the application (in fact not all the classes are loaded into memory together in the first place, they themselves are loaded into memory as and when required) which is a running spring boot ML model microservice at this point. The port at which the microservice exposes its endpoints is also configurable via the application.properties file.

\section {Acumos Model Runner APIs}
In order to understand exactly what functionality, the model runner has provided to the running microservice, it is worthwhile and fitting to look at the APIs of the running microservice (the microservice that was created using our model).

\begin{table*}[hbtp]
\caption{Prominent API endpoints provided by Acumos Model Runner}
\centering
\begin{center}
\begin{tabular}{|p{0.02\linewidth} | p{0.2\linewidth} | p{0.1\linewidth}  |p{0.1\linewidth} |p{0.3\linewidth} |p{0.1\linewidth}|}
\hline
\textbf{} & \textbf{API endpoint} & \textbf{Inputs} & \textbf{Optional} & \textbf{Function} & \textbf{Returns} \\
\hline
1	& /getBinary &	csv data,
proto file	 & operation	& Serializes the csv file based on the .proto file provided here. The .proto file will not replace the default .protofile	& serialized csv \\
\hline
2	& /getBinaryDefault	& csv data	& operation	& Serializes the csv file based on default.proto file.	& serialized csv   \\
\hline
3	& /getBinaryJSON	& json data file,
proto file	& operation	& Serializes the json file based on the .proto file provided here. The .proto file will not replace the default .protofile	& serialized json \\
\hline
4	& /getBinaryJSONDefault	& json data
	&  operation	&  Serializes the json file based on default.proto file.	& serialized json \\
\hline
5	& /model	& ML model (format as specified for the various types of models) &	operation	& Replaces the model with a new model &	- \\
\hline
6	& /model/configuration	& modelConfig file	& operation	& Replaces the current modelConfig file used by Java based models &	- \\
\hline
7	& /proto &	proto file	& operation	& Replaces the default .protofile with the supplied .protofile &	- \\
\hline
8	& /transformCSV	&csv data,
proto file,
ML model	& operation	& Calculates predictions for the csv data, .protofile and ML model combination	& serialized predictions \\
\hline
9	& /transformCSVDefault &	csv data &	operation	& Calculates predictions for the csv data, for the default .protofile and default i.e. current model	& serialized predictions \\
\hline
10	& /transformJSON	& json data,
proto file,
ML model &	operation	& Calculates predictions for the json data, .protofile and ML model combination	& serialized predictions \\
\hline
11	& /transformJSONDefault &	json data file	& operation	& Calculates predictions for the json data, for the default .protofile and default i.e. current model	& serialized predictions \\
\hline

\end{tabular}

\label{table1}
\end{center}
\end{table*}

\section{What can be done using the Model Runner APIs}
Looking at the API endpoints exposed by the model microservice, it is clear that we can do lots of powerful things. Let us outline a few scenarios of how we can leverage the above endpoints to achieve reusable infrastructure i.e., deployment and operations (i.e., serving) for AI / ML models while still having continuously trained AI/ML models in production –

\begin{itemize}
\item We could have an ML model running as a microservice (let is call this the productionalized model) and yet be training a new ML model based on fresh data/ more data at a location wherever we usually train ML models. Once the ML model training is complete, and we have a newer version of the current model, we could call the /model endpoint to replace the current model with the new model. When the call returns, the new model has been productionalized. The new model could be a newer version of the productionalized model or a better performing model.

We have thus accomplished an almost hot swapping of an older version model with a newer ML model. And due to Acumos’ inbuilt functionalities, we save and version the model jars and other artifacts as well as the docker (microservice) images in protected and securely accessed artifact and image repositories.

\item With the powerful API capabilities provided to the microservice by the Generic Model runner, we could go one step further and try to replace the model with a differently behaving newer model (which has different inputs and/or different outputs and/or behaves differently compared to the older existing operationalized model).

A host of different endpoints that allow you to change the behavior of an existing model by replacing it with a new model on the fly without bringing down the microservice are available. If the inputs and/or outputs of the model have changed slightly, you could upload a new proto file at the /proto endpoint. But if the model has changed its behavior: then you want to replace the ML model for e.g., that predicts failures in the 5G mobility network with a new model that predicts what might be the root cause of a known/seen failure. In this case, we can call for e.g., the /transformCSV endpoint which allows us to pass a proto file, the new ML model, and a sample csv that shows how the test or training data looks like.

This changes the behavior of the microservice, so that it can now answer business questions based on the new ML model – this means that the new ML model is now in production. You could then for e.g., call the /getBinaryDefault endpoint with a csv file of test data to get predictions off the standing microservice.

\item The same setup can be used in one more way. Let’s assume you don’t need the model in production anymore. The standing microservice and all the resources associated with it can be repurposed and reused for a new model that is completely different/has nothing to do with the current model in production.

As an example, consider that we do not need the current productionalized ML model i.e., the one that predicts the root cause of failure in the 5G mobility network. We can repurpose and reuse the same standing microservice and also all of the resources we have spent (compute, time, manual effort) to serve a completely different model that for e.g., identifies anomalous IoT devices, or a model that estimates the quality of transmission (QoT) for Optical routing inside Software Defined Optical Networks \cite{shariati2021inter} or a model that predicts which customers will churn this month. The models are entirely different from the model in production. But with the functionality provided by the Acumos Model Runner we can hot swap one model with a completely different ML model without bringing down the microservice itself.

\end{itemize}

\section{Conclusion}
We have seen how the Acumos Model Runner helps us with novel capabilities to reuse the deployment and the infrastructure while operationalizing ML models. This also helps save a lot of time, manual effort, and compute resources, and enables the model change to be made without much service downtime. It also allows us to continuously productionalize latest versions of the model trained on fresh data or more data, or change model behavior on the fly or even change entire models and hence service behaviors on the fly. Reusable MLOps is a novel concept implemented by some Acumos components. It is definitely a step towards the goal of sustainable AI.

\bibliographystyle{IEEEtran}
\bibliography{MyReferences}

\end{document}